\documentclass[nofootinbib,preprintnumbers]{revtex4-2}
\usepackage{amsmath,amssymb,graphicx}
\usepackage{ mathrsfs }
\usepackage{algorithm}
\usepackage{algpseudocode}
\usepackage{xr-hyper}
\usepackage[bottom]{footmisc}
\usepackage{xcolor}
\definecolor{darkred}{rgb}{0.801, 0.0, 0.0}
\usepackage[colorlinks=true, urlcolor=darkred,citecolor=darkred,anchorcolor=darkred,linkcolor=darkred]{hyperref}
\usepackage{ulem}

\def\eqref#1{Eq.$\,$\textcolor{darkred}{(\ref{#1})}}
\def\figref#1{Fig.$\,$\textcolor{darkred}{(\ref{#1})}}
\def\tableref#1{Table$\,$\textcolor{darkred}{(\ref{#1})}}
\def\tr#1{\text{tr}\big(#1\big)}
\def\dtau{\Delta \tau}
\def\dt{\Delta t}
\def\O#1{\mathcal{O}(#1)}
\def\tmax{t_{\text{max}}}
\newcommand{\dbar}{d\hspace*{-0.08em}\bar{}\hspace*{0.1em}}

\begin{document}
\preprint{MIT-CTP/5642}

\title{Real-time Spin Systems from Lattice Field Theory}

\author{Neill~C.~Warrington}
\affiliation{Center for Theoretical Physics, Massachusetts Institute of Technology, Cambridge, MA 02139, USA}

\date{\today}

\begin{abstract}
We construct a lattice field theory method for computing the real-time dynamics of spin systems in a thermal bath. This is done by building on previous work of Takano with Schwinger-Keldysh and functional differentiation techniques. We derive a 
Schwinger-Keldysh path integral for generic spin Hamiltonians, then demonstrate the method on a simple system. Our path integral has a sign problem, which generally requires exponential run time in the system size, but requires only linear storage. The latter may place this method at an advantage over exact diagonalization, which is exponential in both. Our path integral is amenable to contour deformations, a technique for reducing sign problems.
\end{abstract}

\maketitle

\section{Introduction}
In this paper we develop a lattice field theory method for computing the real-time dynamics of  spin systems in a thermal bath. By ``spin-system", we mean a collection of $SU(2)$ generators, $s(x)$, assigned to sites on a lattice, interacting through a Hamiltonian $H(s)$. This lattice could correspond to an actual lattice in space, but may be also be an arbitrary graph. No assumptions are made about the Hamiltonian except that it is a function of a finite number of spins. 

Our purpose is twofold. First, we wish to provide an alternative to exact diagonalization for reproducing the real-time dynamics of digital quantum computers. Today, the only way to check a digital quantum computer's simulation of real-time physics for a general hamiltonian is to diagonalize it or perform a Trotterized task of equal complexity \cite{Qiskit,PhysRevLett.121.170501,PhysRevA.99.052335,PhysRevLett.123.090501}. The cost of such computations is exponential in the number of qubits in both storage and run-time. The method we develop is exponential in runtime, but is linear in storage. Second, we wish to provide an alternative to tensor networks for computing the real-time properties of spin systems, especially in more than one spatial dimension, where tensor networks are generally less efficient \cite{orus,Orus:2018dya}.

Our main result is a path integral discretization for spin systems which provably reproduces the exact real-time dynamics for any Hamiltonian of a finite number of spins. The path integral is discretized in both space and time, and the desired dynamics are recovered in the  ``time continuum limit". The field variables of this path integral are points on the two-sphere which are in one-to-one correspondance with points on the Bloch sphere. What is shown is that correlation functions of points on the Bloch sphere in the lattice theory converge exactly to spin correlation functions in the quantum theory. What is perhaps unconventional about our formulation is the appearance of a ``Schwinger-Keldysh action" \cite{schwinger,Keldysh:1964ud}, which encodes both real-time and thermal information. 

The most important property of our discretization is that has the correct continuum limit. The same cannot be said of all  discretizations on the market. Demonstrated analytically in \cite{wilson-galitski} then corroborated numerically in \cite{PhysRevB.106.214416}, even textbook path integrals for spin systems, such as those found in \cite{altland-and-simons,assa}, do not in general have time continuum limits. In spite of being explicitly mentioned \cite{assa}, this fact appears little appreciated today.

Our discretization builds upon a foundation layed by Takano in the 1980's \cite{takano}, where he developed a Monte Carlo method for computing thermal properties of spin systems. The basic building block of Takano's path integral are ``spin coherent states"\footnote{In the literature these are also called ``Bloch coherent states" (e.g. in \cite{arecchi}).}. Developed in \cite{radcliffe,atkins,kutzner,arecchi}, spin coherent states are to spin operators what bosonic coherent states are to bosonic operators: they are an overcomplete set of states providing a resolution of the identity that can be used to build a path integral. The resolution of the identity involves an integral over the Bloch sphere, and this is ultimately why the path integral has field variables on the sphere.


What we add to Takano's work is real-time dynamics and simplification. In his original work the possibility of real-time dynamics was not  considered. Furthermore, the original formulation relies on a certain ``product formula" for inserting observables into the thermal trace \cite{takahashi-and-shibata}. Implementing the product formula quickly becomes prohibitively difficult as both both the hamiltonian and observables increase in complexity. We skirt the product formula altogether by using functional differentiation, a common technique in lattice field theory for observable insertions. This allows for the analysis of more general Hamiltonians and observables than would have otherwise been practical. In the following sections we derive the path integral, demonstrate that it works, then offer a discussion.

\section{Derivation}
\label{formalism}
Consider a spatial lattice of spins $s(x)$ interacting through a hamiltonian $H$ and satisfying the canonical commutation relations $[s_i(x),s_j(y)] = i\epsilon_{i j k}s_k(x) \delta_{xy}$. In this section we develop a path integral representation of correlation functions
\begin{equation}
\label{eq:corr-fcns}
    \langle s_i(x,t) s_{i'}(x',t') \rangle \equiv \frac{\tr{s_i(x,t) s_{i'}(x',t') e^{-\beta H}}}{\tr{e^{-\beta H}}}
\end{equation}
where $x,x'$ are lattice sites, $t,t'$ are real times, and $\beta =T^{-1}$ is the inverse temperature of the system. This path integal representation is obtained in two steps. First, a trotterized, Hilbert-space object is formed which reproduces the correlation functions in the time continuum limit. Second, this Hilbert-space object is approximated by a path integral with controlled errors. While our presentation is for two-point functions it will be clear the procedure generalizes to n-point functions. Throughout we assume a lattice of finite extent but arbitrary spatial dimension.

\begin{figure}[b]
    \centering
    \includegraphics[height=2.5in]{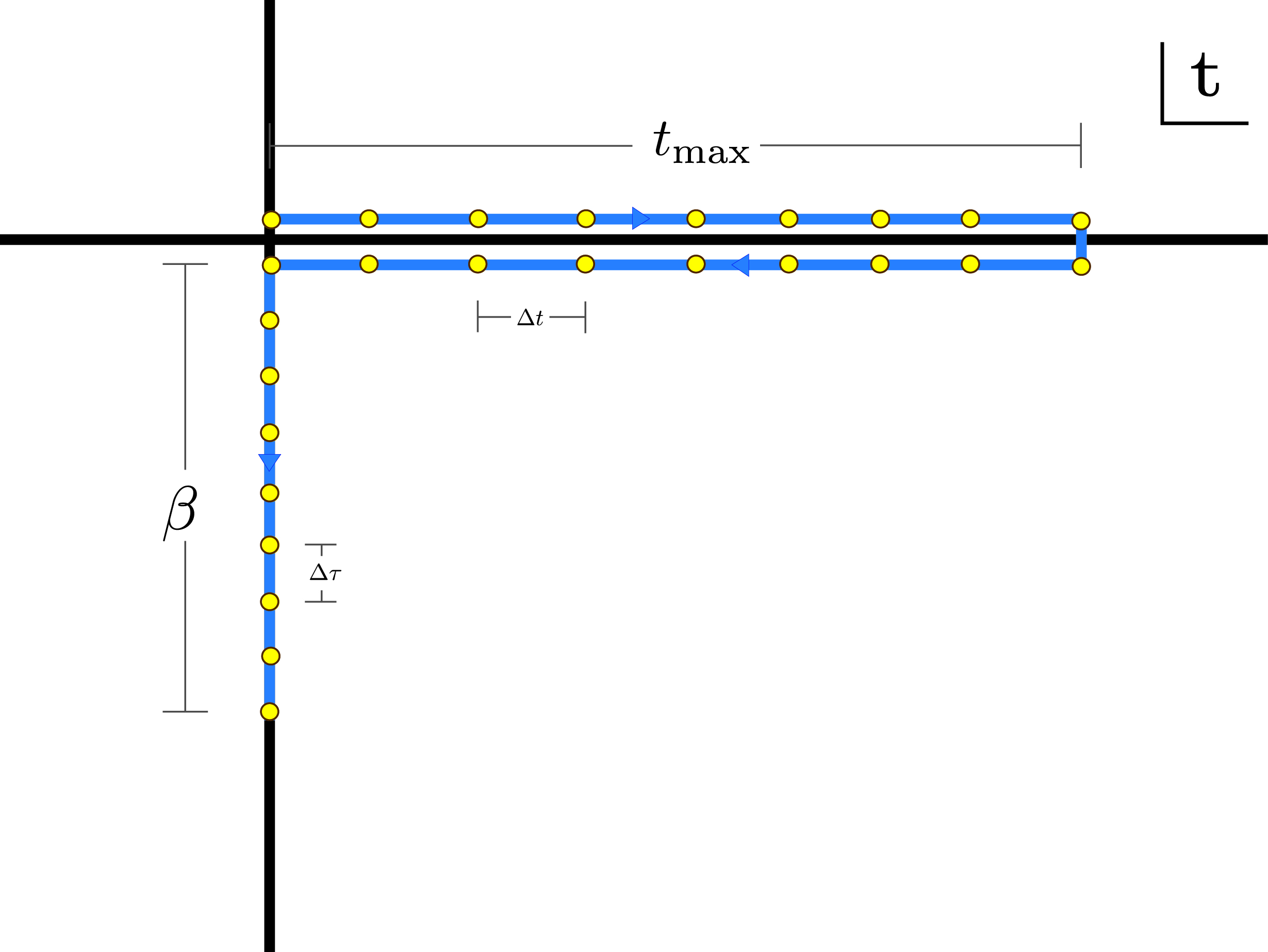}
    \caption{Schematic of the discretized Schwinger-Keldysh contour. The path followed is indicated by the arrows: beginning at zero the contour sweeps right to positive $t_{\text{max}}$, then back to zero, then down an amount $\beta$. The vertical splitting between the real-time segments is only drawn so that the forward and backward contours do not overlap; an $i \epsilon$ prescription is neither needed nor employed. }
    \label{fig:sk-contour}
\end{figure}

Consider a time extent $\tmax$ larger than $t$ $\&$ $t'$ and the following object:
\begin{align}
\label{eq:Ztilde}
    \widetilde{Z}(j^+, j^-, j^E) = 
    \text{tr} \Bigg( \prod_{t=1}^N \widetilde{P}_+(j^{+}_t) \prod_{t=1}^N \widetilde{P}_-(j^{-}_t) \prod_{t=1}^N \widetilde{P}_E(j^{E}_t)  \Bigg) \, ,
\end{align}
where
\begin{align}
    \widetilde{P}_{+}(j^+_t) & = \mathbf{1}+ i \dt \big(H - \sum_x j^+_t(x)\cdot s(x)\big) \nonumber \\
    \widetilde{P}_{-}(j^-_t) & = \mathbf{1}- i \dt \big(H - \sum_x 
    j^-_t(x)\cdot s(x)\big) \nonumber \\
    \widetilde{P}_{E}(j^E_t) & = \mathbf{1}- \dtau \big(H - \sum_x j^+_E(x)\cdot s(x)\big)~,
\end{align}
where the sums run over lattice sites, $\mathbf{1}$ is the identity operator, $N$ is an integer number of ``timeslices", $\dtau=\beta/N$, $\dt=\tmax/N$, and $j^+,j^-, j^E$ are classical sources. For brevity, we collect all sources into a symbol $j$ and define $\widetilde{Z}(j) \equiv \widetilde{Z}(j^+, j^-, j^E)$. 
The short-time propagators  $\widetilde{P}_{+},\widetilde{P}_-,\widetilde{P}_E$ build up a discretized ``Schwinger-Keldysh contour" in the complex time plane, shown in \figref{fig:sk-contour}. This contour defines a ``Schwinger-Keldysh action" \cite{schwinger,Keldysh:1964ud}, which we will elaborate on later.

We presently show that derivates of $\widetilde{Z}$ with respect to sources produce the real-time correlation functions \eqref{eq:corr-fcns} when $N\rightarrow \infty$ at fixed $\beta, t_{\text{max}}$. We borrow language from lattice field theory and call this the ``time-continuum limit". First, since the Hilbert space is finite dimensional, the Hamiltonian is bounded and $N$ can always be taken large enough that
\begin{align*}
    \widetilde{P}_E(0) & = e^{-\dtau H} + \O{\dtau^2} \\
    \widetilde{P}_{\pm}(0) & = e^{\pm i \dt H} + \O{\dt^2} \, ,
\end{align*}
where the errors are small as desired. Therefore $\widetilde{Z}(0)$ converges linearly to $\tr{e^{-\beta H}}$ in the time continuum limit. Next, consider two spacetime points $(x,t)$ $\&$ $(x',t')$ with $t<t'<t_{\text{max}}$,
and define integers $(\hat{t},\hat{t}', \hat{t}_{\text{max}}) = (t, t',t_{\text{max}})/\Delta t$. Then
\begin{equation}
    \Big( \frac{1}{-i \dt} \Big)^2 ~ \frac{1}{\widetilde{Z}(j)} ~ \frac{\partial}{\partial j^+_{\hat{t},i}(x)}
    \frac{\partial}{\partial j^+_{\hat{t}',i'}(x')}\widetilde{Z}(j)\Bigg \rvert_{j = 0}  = \frac{1}{\widetilde{Z}(0)} \text{tr} \Bigg( \widetilde{P}_+^{\hat{t}-1} s_i(x) \widetilde{P}_+^{\hat{t}'-\hat{t}-1} s_{i'}(x') \widetilde{P}_+^{N-\hat{t}'} \widetilde{P}_-^{N} \widetilde{P}_E^N\Bigg)
\end{equation}
where $\widetilde{P}_{a} \equiv \widetilde{P}_{a}(0)$ for $a=+,-,E$. Since $\widetilde{P}_+^{\hat{t}-1} = e^{i H t} + \mathcal{O}(\dt)$, and similarly for the other products of short time propagators, one obtains 
\begin{equation}
    \Big( \frac{1}{i \dt} \Big)^2 ~ \frac{1}{\widetilde{Z}(j)} ~ \frac{\partial}{\partial j^+_{\hat{t},i}(x)}
    \frac{\partial}{\partial j^+_{\hat{t}',i'}(x')}\widetilde{Z}(j)\Bigg \rvert_{j = 0}  
    = \langle s_i(x,t) s_{i'}(x',t' ) \rangle + \mathcal{O}(\dt, \dtau),  \text{ for } t< t'.
\end{equation}
Taking a time-continuum limit while holding the physical time separation between derivatives fixed, a real-time correlator is obtained. While this construction produced a \textit{time-anti-ordered} correlator, any time ordering can be obtained. One finds:
\begin{align}
    \Big( \frac{1}{i \dt} \Big)^2 ~ \frac{1}{\widetilde{Z}(j)} ~ \frac{\partial}{\partial j^-_{\hat{t},i}(x)}
    \frac{\partial}{\partial j^-_{\hat{t}',i'}(x')}\widetilde{Z}(j)\Bigg \rvert_{j = 0}  
    & = \langle s_i(x,t) s_{i'}(x',t' ) \rangle + \mathcal{O}(\dt, \dtau),  \text{ for } t>t', \nonumber \\
    -\Big( \frac{1}{i \dt} \Big)^2 ~ \frac{1}{\widetilde{Z}(j)} ~ \frac{\partial}{\partial j^+_{\hat{t},i}(x)}
    \frac{\partial}{\partial j^-_{\hat{t}_{\text{max}} - \hat{t}',i'}(x')}\widetilde{Z}(j)\Bigg \rvert_{j = 0}  
    & = \langle s_i(x,t) s_{i'}(x',t' ) \rangle + \mathcal{O}(\dt, \dtau),  \text{ for any } t, t' .
\end{align}

We will now relate $\widetilde{Z}(j)$ to a lattice path integral using Takano's framework \cite{takano}. The key step is to express the short-time propagators in terms of spin coherent states, which is accomplished using a ``diagonalization theorem" \cite{arecchi,radcliffe,lieb}. We first set notation: to every spatial lattice site $x$ we associate a sphere whose points we denote $\Omega_x = \begin{pmatrix}\text{sin}\theta_x \text{cos}\phi_x, & \text{sin}\theta_x \text{sin}\phi_x, & \text{cos}\theta_x  \end{pmatrix}$. A spin coherent state is then defined as 
\begin{equation}
    | \Omega \rangle = \prod_x |\Omega_x \rangle \equiv \prod_x e^{i\theta_x(s_1 \text{sin}\phi_x - s_2 \text{cos} \phi_x)}| s, s \rangle~,
\end{equation}
where $|s,s\rangle$ is the highest weight state in the spin-$s$ representation of $SU(2)$. Then the diagonalization theorem states that any operator on spin space may be written as an integral over spin coherent states:
\begin{equation}
\label{eq:diag-theorem}
    \mathcal{O} = \int \dbar \Omega ~ f_{\mathcal{O}}(\Omega) ~ | \Omega \rangle \langle \Omega |~,
\end{equation}
where the function $f_{\mathcal{O}}$ depends on the choice of operator and $\dbar \Omega = \prod_x \frac{(2s+1)}{4\pi} d\Omega_x $. Useful examples include
\begin{align}
    \mathbf{1} & = \int \dbar \Omega ~ 1 ~ | \Omega \rangle \langle \Omega | \nonumber \\
    s(x) & = \int \dbar \Omega ~ (s+1) \Omega_x ~ | \Omega \rangle \langle \Omega |  ~.
\end{align}
The theorem allows to write
\begin{equation}
    \widetilde{P}_+(j^+_t) = \int \dbar \Omega \Bigg(1+i \dt \Big[ h(\Omega) - (s+1) \sum_x \Omega_x \cdot j^+_t(x) \Big] \Bigg) ~ | \Omega \rangle \langle \Omega | 
\end{equation}
and similarly for $\widetilde{P}_-,\widetilde{P}_E$\footnote{Since the only way the spin representation appears in the coherent state picture is through the constant $s$, it is simple generalize our procedure to site-dependent spin representations.}. Here $h$ is related to the Hamiltonian through $H = \int \dbar \Omega ~ h(\Omega) ~ | \Omega \rangle \langle \Omega |$. We now introduce the operators\footnote{Note that $\Omega^{\pm},\Omega^E$ are simply suggestive variable names: all three are points on the sphere.}
\begin{align}
\label{eq:the-ps}
    P_+(j^+_t) & = \int \dbar \Omega^+ e^{+i \dt \big[ h(\Omega^+) - (s+1) \sum_x \Omega^+_x \cdot j^+_t(x) \big]} ~ | \Omega^+ \rangle \langle \Omega^+ | \nonumber \\
    P_-(j^-_t) & = \int \dbar \Omega^- e^{-i \dt \big[ h(\Omega^-) - (s+1) \sum_x \Omega^-_x \cdot j^-_t(x) \big]} ~ | \Omega^- \rangle \langle \Omega^- | \nonumber \\
    P_E(j^E_t) & = \int \dbar \Omega^E e^{-\dtau \big[ h(\Omega^E) - (s+1) \sum_x \Omega^E_x \cdot j^E_t(x) \big]} ~ | \Omega^E \rangle \langle \Omega^E |~,
\end{align}
which are equal to $\widetilde{P}_+(j^+_t),\widetilde{P}_-(j^-_t),\widetilde{P}_E(j^E_t)$ up to errors of $\mathcal{O}(\dt^2, \dtau^2)$. Therefore, 
\begin{align}
\label{eq:Z}
    Z(j) = 
    \text{tr} \Bigg( \prod_{t=1}^N P_+(j^{+}_t) \prod_{t=1}^N P_-(j^{-}_t) \prod_{t=1}^N P_E(j^{E}_t)  \Bigg) \, ,
\end{align}
is equal to $\widetilde{Z}(j)$ up to errors of $\mathcal{O}(\dt,\dtau)$. Computing the trace, one finds:
\begin{equation}
\label{eq:SK-path-integral}
    Z(j) = \int D\Omega
    ~ e^{S_{SK}(\Omega,j)}~,
\end{equation}
where:
\begin{align}
\label{eq:SK-definitions}
    e^{S_{SK}(\Omega,j)} = &  
    ~ e^{
    i \dt \sum\limits_{ \hat{t} } \Big(h(\Omega^+_{ \hat{t} }) -(s+1) \sum\limits_x \Omega^+_{\hat{t}\, x} \cdot j^+_t(x)\Big)
    -i \dt \sum\limits_{\hat{t}} \Big(h(\Omega^-_{ \hat{t} }) -(s+1) \sum\limits_x \Omega^-_{\hat{t} \, x} \cdot j^-_t(x)\Big)
    - \dtau \sum\limits_{ \hat{t} } \Big(h(\Omega^E_{ \hat{t} }) -(s+1) \sum\limits_x \Omega^E_{\hat{t} \, x} \cdot j^E_t(x)\Big)
    } \nonumber \\
    & \times \prod_t \langle \Omega_t | \Omega_{t+1} \rangle \nonumber \\
    \prod_t \langle \Omega_t | \Omega_{t+1} \rangle = & ~ \langle \Omega^+_1 | \Omega^+_2 \rangle ... \langle \Omega^+_N | \Omega^-_1 \rangle \langle \Omega^-_1 | \Omega^-_2 \rangle ... \langle \Omega^-_N | \Omega^E_1 \rangle \langle \Omega^E_1 | \Omega^E_2 \rangle ... \langle \Omega^E_N | \Omega^+_1 \rangle \nonumber \\
    D\Omega = & \prod_{t} \dbar \Omega^+_t \prod_{t} \dbar \Omega^-_t \prod_{t} \dbar \Omega^E_t ~.
\end{align}
We call $S_{SK}(\Omega,j)$ the Schwinger-Keldysh action. To clarify the notation, the $\Omega^+_t$ appearing within $Z(j)$ arise from the product of $P_+$ operators, while the $\Omega^-_t$ and $\Omega^E_t$ arise from the products of $P_-$ and $P_E$. We collect these three into a composite variable $\Omega$ which is a ``path ordered" concatenation of $\Omega^+, \Omega^-,\Omega^E$, in that order. Referring to the Schwinger-Keldysh contour in \figref{fig:sk-contour}, $\Omega^+_t,\Omega^-_t, \Omega^E_t$ are assigned to the forward, backward, and Euclidean segments, respectively. Since $\widetilde{Z}(j)$ and $Z(j)$ differ at $\mathcal{O}(\dt,\dtau)$, so do their derivatives. One therefore has the following relations, up to errors of $\mathcal{O}(\dt, \dtau)$, which vanish in the time continuum limit:
\begin{align}
\label{eq:main-results}
    (s+1)^2 \langle \Omega^+_{\hat{t} \, x \, i}  \Omega^+_{\hat{t'} \, x' \, i'} \rangle_{SK} & = \langle s_i(x,t) s_{i'}(x',t' ) \rangle,  \text{ for } t< t' \text{ (time-anti-ordered)} \nonumber \\
    (s+1)^2 \langle \Omega^-_{\hat{t} \, x \, i}  \Omega^-_{\hat{t'} \, x' \, i'} \rangle_{SK} & = \langle s_i(x,t) s_{i'}(x',t' ) \rangle,  \text{ for } t > t' \text{ (time-ordered)} \nonumber \\
    (s+1)^2 \langle \Omega^+_{\hat{t} \, x \, i}  \Omega^-_{\hat{t}_{\text{max}} -\hat{t'} \, x' \, i'} \rangle_{SK} & = \langle s_i(x,t) s_{i'}(x',t' ) \rangle,  \text{ for any } t, t' \text{ (unordered)} ~.
\end{align}
Here $\langle \mathcal{O} \rangle_{SK} \equiv Z(0)^{-1} \int D\Omega ~e^{S_{SK}(\Omega,0)} \mathcal{O}(\Omega)$ and $\Omega^a_{\hat{t} \, x \, i}$ is the $i^{\text{th}}$ component of $\Omega^a$ on lattice site $x$ and timeslice $\hat{t}$. \eqref{eq:SK-definitions} and \eqref{eq:main-results} are our main results. 

We conclude with some details about the Schwinger-Keldysh action. The inner product of coherent states appearing in \eqref{eq:SK-definitions} is the analog of the ``Berry phase" in our formulation \cite{assa}. Each overlap has the explicit expression
\begin{equation}
    \langle \Omega' | \Omega \rangle = \prod_x \Big(\text{cos}\frac{\theta_x'}{2}\text{cos}\frac{\theta_x}{2} + e^{-i(\phi_x' -\phi_x)}\text{sin}\frac{\theta_x'}{2}\text{sin}\frac{\theta_x}{2}\Big)^{2s}~.
\end{equation}
The only dependence of the path integral on the system in question is through the structure of the lattice and the function $h(\Omega)$. As long as the hamiltonian does not involve products of spins on the same site, $h$ is obtained from the Hamiltonian through the simple relation:
\begin{equation}
\label{eq:h-function}
    h(\Omega) = H\big(\hat{s} \rightarrow (s+1) \Omega \big)~,
\end{equation}
where on the right hand side we have distinguished the operator $\hat{s}$ from the spin representation $s$. If the Hamiltonian does contain products of spins on the same site, \eqref{eq:h-function} no longer holds, and it becomes necessary to use the product formula of \cite{takano,takahashi-and-shibata} to derive $h$. We emphasize however that the only complication is in the relation between $H$ and $h$. Correlation functions retain the form \eqref{eq:main-results}.


\section{Demonstration}
\label{application}
We now engage in a numerical demonstration. We consider the Hamiltonian
\begin{equation}
    H = -\frac{j}{2} \sum_x s_1(x) s_1(x+1) + s_3(x) s_3(x+1)~,
\end{equation}
for two-qubits in one spatial dimension with periodic  boundary conditions. We will obtain the real-time correlation functions $\langle s_i(x,t) s_{i'}(x',t') \rangle$ by computing   the Schwinger-Keldysh path integral \eqref{eq:SK-path-integral} and taking a time-continuum-limit. The $h$ function corresponding to this hamiltonian is 
\begin{equation}
     h = - \frac{j(s+1)^2}{2} \sum_x \Omega_1(x) \Omega_1(x+1) + \Omega_3(x) \Omega_3(x+1)~.
\end{equation}
Usually, lattice path integrals can only be estimated stochastically using Monte Carlo methods. However, by considering a small system, we are able to elide the Monte Carlo and compute the lattice path integral exactly. Without the burden of stochastic noise we demonstrate that the time continuum limit of the Schwinger-Keldysh path integral \eqref{eq:SK-path-integral} both exists and converges to the correct result. Since the method works for multiple qubits, albeit two, we see no reason it should fail for any finite number of them. 

To circumvent the Monte Carlo, we compute matrix representations of the $P_+,P_-,P_E$ appearing in \eqref{eq:Z}. For this two-qubit theory, these are $4 \times 4$ matrices which, at fixed $\dt$ and $\dtau$, can be 
computed numerically by evaluating the integrals over spheres in \eqref{eq:the-ps}.  To form correlation functions, we also numerically compute the matrices
\begin{equation}
    \int \dbar \Omega^{\pm} e^{\pm i \dt \, h(\Omega^+) } ~ \Omega^{\pm}_{x \, i} ~ | \Omega^{\pm} \rangle \langle \Omega^{\pm} |~.
\end{equation}
In the following example we take $\beta = 3.0 j^{-1}$ and $t_{\text{max}} = 10 j^{-1}$. Due to spacetime and internal symmetries, there are only four independent two-point correlation functions in this theory: $\langle s_i(x,t) s_i(x,0) \rangle, \langle s_i(x,t) s_i(x+1,0) \rangle$ for $i=x,y$. We show results only for $i=x$ correlation functions;  the qualitative conclusions drawn from $i=y$ are the same. 

\begin{figure*}[b]
    \centering
       \includegraphics[width=3.5in]{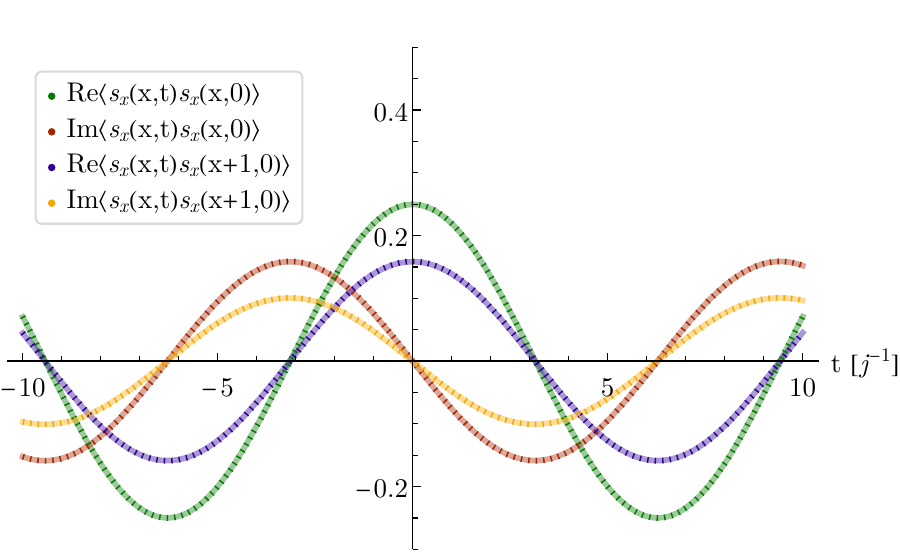}
       \includegraphics[width=3.5in]{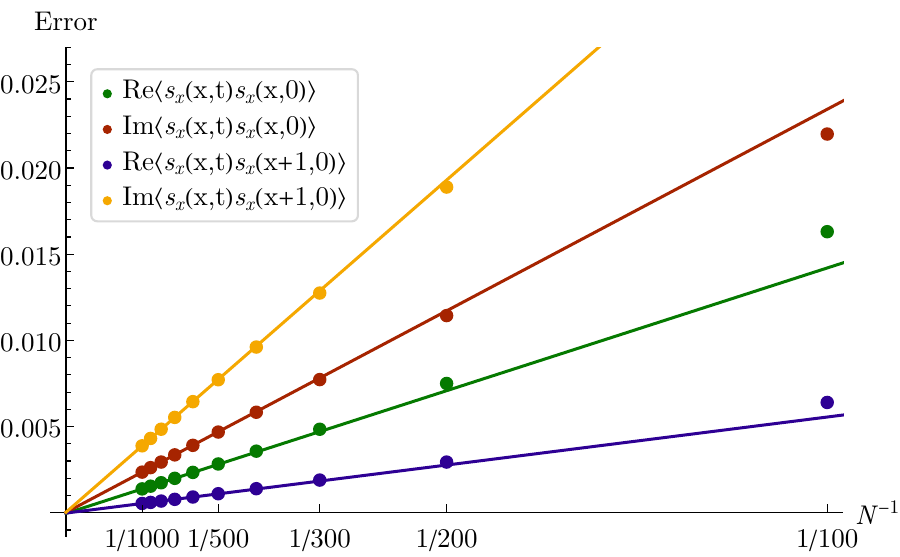}
        \caption{\textbf{Left:} Real-time correlation functions. The transparent curves are unordered correlators obtained from the disctetized theory at fixed $N=5000$, while the dashed curves are the exact result obtained by matrix exponentiation. \textbf{Right:} Time continuum limits at fixed $t = 5.0 j^{-1}$. Error is defined as the difference between the lattice and exact theories, and each point is obtained from a lattice calculation with $N$ timesteps. The lines are linear fits to the finest three lattice spacings.}
\label{fig:results}
\end{figure*}

Our numerical results are found in \figref{fig:results}. In the left panel we plot the exact correlation functions $\langle s_x(x,t) s_x(x',0) \rangle$ in dotted lines, and the unordered correlation functions $(s+1)^2 \langle \Omega^+_{\hat{t} \, x \, i}  \Omega^-_{\hat{t}_{\text{max}} \, x' \, i'} \rangle_{SK}$ in transparent lines. The lattice theory is taken to have $N=5000$ timeslices. The visual agreement between the dashed and transparent curves is evidence the lattice method works. This evidence is corroborated in the right panel of \figref{fig:results}, which illustrates time continuum limits. The vertical axis is the difference between the lattice and exact results at fixed time separation $t=5.0 j^{-1}$. The points are the results of lattice calculations at different number of timeslices $N$, and the lines are linear fits to the finest three lattice spacings. Linear convergence to zero as $N^{-1} \rightarrow 0$ demonstrates both the existence and correctness of a continuum limit of the lattice theory. Clearly the fits in the right panel of \figref{fig:results} converge very close to zero. However, our fits to not converge to $exactly$ zero. This is simply because at any non-zero lattice spacing there are always some $\mathcal{O}(\dt^2,\dtau^2)$ errors  contaminating the fit. We have therefore performed fits at extremely fine lattice spacing. As tabulated in \tableref{table:improving-convergence}, we find a strictly shrinking error as we approach the continuum. We find this to be evidence beyond reasonable doubt that the method works. While we have presented detailed results only for the $\langle \Omega^+  \Omega^-\rangle_{SK}$ correlator, which is able to access twice the time separation as the $\langle \Omega^+  \Omega^+\rangle_{SK}$ and $\langle \Omega^-  \Omega^-\rangle_{SK}$ correlators, we have computed all three and find they all converge to the correct continuum, limit. We have also computed the $\langle s_y s_y \rangle$ correlation functions and obtain correct continuum limits in all cases.

\begin{table}[t!]
\centering
\begin{tabular}{||c | c | c | c | c ||} 
 \hline
  $\text{Fit window}$ & $\text{Re}\langle s_x(x,t)s_x(x,0)\rangle$ & $\text{Im}\langle s_x(x,t)s_x(x,0)\rangle$ & $\text{Re}\langle s_x(x,t)s_x(x+1,0)\rangle$ & $\text{Im}\langle s_x(x,t)s_x(x+1,0)\rangle$ \\ [0.5ex] 
  \hline
   $\{300,400,500\}$       & $-2.0 \times 10^{-4}$ & $1.3 \times 10^{-4}$ & $-8.2 \times 10^{-5}$ & $1.8 \times 10^{-4}$ \\
   \hline
   $\{1000,1500,2000\}$    & $-1.6 \times 10^{-5}$ & $1.1 \times 10^{-5}$ & $-6.9 \times 10^{-6}$ & $1.5 \times 10^{-5}$ \\
   \hline
   $\{3000,4000,5000\}$    & $-2.3 \times 10^{-6}$ & $1.4 \times 10^{-6}$ & $-1.0 \times 10^{-6}$ & $1.9 \times 10^{-6}$ \\
   \hline
   $\{10000,12500,15000\}$ & $-4.0 \times 10^{-7}$ & $2.5 \times 10^{-7}$ & $-2.2 \times 10^{-7}$ & $2.6 \times 10^{-7}$ \\
 \hline
\end{tabular}
\caption{Table of continuum extrapolation errors. Printed within each row is the error in a linear continnum extrapolation of three lattice calculations with the number of timeslices indicated by the fit window. The column labels indicate the observable whose error is printed. }
\label{table:improving-convergence}
\end{table}

\section{Discussion}
\label{conclusions}
In this paper we developed a lattice field theory method for computing the real-time dynamics of spin systems. This method extends previous work by Takano with Schwinger-Keldysh and functional differentiation techniques. Our main results are \eqref{eq:SK-definitions} and \eqref{eq:main-results}, which respectively give the Schwinger-Keldysh action for generic hamiltonians and formulae for spin correlation functions. We then demonstrated the method on a simple two-qubit system.

Our method can be easily extended in several ways. First, convergence to the continuum can be accelerated with higher-order approximations to the short-time propagators. The cost of this acceleration is the need to write products like $H^2,H^3,...$ in terms of coherent states which requires the product formula. Such expressions are cumbersome though not impossible to write. Second, three-point and higher-order spin correlation function are simple to obtain by taking more functional derivatives. In fact, n-point functions of arbitrary observables $\mathcal{O}$ are simple to obtain too. From the beginning one simply couples sources to $\mathcal{O}$ rather than to spins. The same Schwinger-Keldysh path integral results, but rather than $\Omega$ correlators one requires $f_{\mathcal{O}}$ correlators ($f_{\mathcal{O}}$ is defined through \eqref{eq:diag-theorem}). Combining this fact with the Schwinger-Keldysh path integral's ability to compute any time ordering, ``out of time ordered correlators" of arbitrary observables can be computed with our method. This may prove to be useful in studies of quantum chaos \cite{Garcia-Mata:2022voo,PhysRevB.97.144304}. Third, though we have chosen a particular approach to the continuum (i.e. we have taken the number of timeslices to be idential on all three legs of the Schwinger-Keldysh contour), infinity other approaches exist and may be more efficient. Indeed, in highly asymmetric cases such as high-temperate $\&$ long real-time, or low-temperature $\&$ short real-time, it may be profitable to choose a different number of timeslices on each of the legs. Finally, we emphasize that no ``$+i \epsilon$" prescription ever appears in our method. It is not needed. This is important: it is unwise to add unnecessary parameters to lattice simulations requiring extrapolation. There are already enough to do.

A Monte Carlo evaluation of the path integral will be unavoidable for larger systems. While the storage cost for Monte Carlo evaluations of lattice path integrals is linear in the number of space time points, our path integral has a sign problem. This generally requires exponential runtime to resolve \cite{troyer-wiese}. However, our path integral has a holomorphic path integrand for any Hamiltonian. This means it is anemable to ``path integral contour deformations", a general method for taming sign problems where the integration manifold is deformed into complex field space to a surface with reduced phase oscillations \cite{RevModPhys.94.015006}. That our path integral is over many spheres makes it particularly straightforward to apply contour deformations, and existing machine-learned manifolds developed for other systems may prove useful \cite{PhysRevD.103.094517}. While the real-time dynamics of spins is a challenging problem, it is encouraging that other real-time problems have been solved with contour deformations \cite{PhysRevLett.117.081602,PhysRevD.95.114501}.

\section{Acknowledgements}
We are grateful to Hank Lamm, Phiala Shanahan, and Paulo Bedaque for their time and discussions. NCW is supported in part by the U.S. Department of Energy, Office of Science under grant Contract Numbers DE-SC0011090 and DE-SC0021006 and by Simons Foundation grant 994314 (Simons Collaboration on Confinement and QCD Strings)

\bibliography{bibliography} 
\end{document}